\documentclass[manuscript]{aastex}
\usepackage{spr-astr-addons}

\shorttitle{Wormholes}

\shortauthors{Rahaman et al.}

\begin{document}

\title{Could wormholes form in dark matter galactic halos?}

\author{Farook Rahaman\altaffilmark{1}}
\altaffiltext{1}{Department of Mathematics, Jadavpur University,
Kolkata 700032, West Bengal, India\\rahaman@iucaa.ernet.in}

\author{G.C. Shit\altaffilmark{2}}
\altaffiltext{2}{Department of Mathematics, Jadavpur University,
Kolkata 700 032, West Bengal, India\\gopal\_iitkgp@yahoo.co.in}

\author{Banashree Sen\altaffilmark{3}}
\altaffiltext{3}{Department of Mathematics, Jadavpur University,
Kolkata 700 032, West Bengal, India\\banashreesen7@gmail.com}

\author{Saibal Ray\altaffilmark{4}}
\altaffiltext{4}{Department of Physics, Government College of
Engineering \& Ceramic Technology, Kolkata 700010, West Bengal,
India\\saibal@iucaa.ernet.in}

\begin{abstract} We estimate expression for velocity as a function of the
radial coordinate $r$ by using polynomial interpolation based on
the experimental data of rotational velocities at distant outer
regions of galaxies. The interpolation technique has been used to
estimate fifth degree polynomial followed by cubic spline
interpolation. This rotational velocity is used to find the
geometry of galactic halo regions within the framework of
Einstein's general relativity. In this letter we have analyzed
features of galactic halo regions based on two possible choices
for the dark matter density profile, viz. Navarro, Frenk \& White
(NFW) type~\citep{Navarro1996} and Universal Rotation Curve (URC)
~\citep{Castignani2012}. It is argued that spacetime of the
galactic halo possesses some of the characteristics needed to
support traversable wormholes.
\end{abstract}


\section{Introduction}

During last several decades wormhole has been attracting attention
to the scientific community a lot after since publication of the
seminal work by~\citet{Morris1988}. In this paper they argued the
possibility of the existence of traversable wormholes permitting
to travel through space and time. Actually, a wormhole does act
role for a passage/tunnel in spacetime which is supposed to
connect the widely separated regions of our universe or different
universes in the multiverse model. According to
~\citet{Morris1988}, the normal matter is unable to hold a
wormhole open rather the matter is responsible for sustaining a
traversable wormhole is exotic in nature which violates the
standard null energy condition.

By drawing our attention to the galactic level where one can
notice a kind of peculiar phenomena known as the flat rotation
curves in galaxies. Now, the ordinary luminous matters in space
are composed mainly of neutral hydrogen clouds. But this bizarre
galactic rotation curves can not be explained by the standard
model. Therefore, to explain the rotation curves in the outer
regions of galaxies it has been supposed that galaxies and even
clusters of galaxies must contain some non-luminous matter. This
kind of exotic stuff, now known as dark matter, does not emit
electromagnetic waves nor interact with normal matter. It has
arguably proved by the scientists that dark matter can explain
properly the so called flat rotation curves.

In the previous some studies in connection to flat rotation
curves~\citep{Rahaman2014a,Rahaman2014b} we have used {\it ad hoc}
functional forms of rotation velocity. However, by using the
experimental data of the rotational velocities at different
distance of outer regions of galaxies, we have estimated fifth
degree polynomial that yields the expression for the velocity as a
function of the radial coordinate $r$ which is almost of the same
nature of the experimental feature. Thus in this paper, we study
the exact observational results and confirm the existence of
wormhole in all the galaxies containing the dark matter. These
dark matters actually play the role of fuel for developing
wormhole-like geometry. This study is a combination of Einstein's
theory and experimental result, therefore, reasonably more
physical than the previous
studies~\citep{Rahaman2014a,Rahaman2014b}.

\section{The basic equations and their solutions}

To connect the dark matter and galactic rotation curves with
wormholes one needs to introduce the metric for a static
spherically symmetric spacetime. Following~\citet{Morris1988} we
consider here the line element for the galactic halo region in
connection to wormhole spacetime as
\begin{equation}\label{E:line1}
ds^2=-e^{2f(r)}dt^2+\left(1-\frac{b(r)}{r}\right)^{-1}dr^2
+r^2(d\theta^2+\sin^2\theta\,d\phi^2),
\end{equation}
where $c=G=1$ in the geometrized unit.

Till now we have some idea regarding the density profile, however,
other properties of dark matter yet to be explored. Therefore, it
is justified to assume that dark matter is characterized by the
general anisotropic energy-momentum tensor~\citep{Bozorgnia2013}
\begin{equation}
 T_\nu^\mu=(\rho + p_t)u^{\mu}u_{\nu} - p_t g^{\mu}_{\nu}+
            (p_r -p_t )\eta^{\mu}\eta_{\nu},
\end{equation}
where $u^{\mu}u_{\mu} = - \eta^{\mu}\eta_{\mu} = -1$. Here $p_t$
and $p_r$ are transverse and radial pressures, respectively.

Using Navarro-Frenk-White (NFW)~\citep{Navarro1996} density
profile and constant rotational velocity $v_\phi$,
~\citet{Rahaman2014a} has demonstrated the possible existence of
wormholes in the outer regions of the galactic halo. In another
study,~\citet{Rahaman2014b} have used the Universal Rotation Curve
(URC) dark matter model and sample rotation curve utilizing an
ansatz for $v_\phi$ to obtain analogous results for the central
parts of the halo. This letter is a significant sequel of the
earlier results in a more refined manner and therefore confirms
the possible existence of wormholes in most of the galaxies.

Basically by employing the experimental data of the rotational
velocities of the outer regions of galaxies, we have estimated
fifth degree polynomial that yields the expression for the
velocity as a function of the radial coordinate $r$. The
interpolation technique has been used to estimate this fifth
degree polynomial. This rotational velocity is used to find the
geometry of galactic halo regions within the framework of general
theory of relativity.

As a background of the present study we would like to report here
briefly about the necessary density profiles in the following two
paragraphs.

\subsection{The Navarro-Frenk-White density profile}

\citet{Navarro1996} proposed from the predictions of standard cold
dark matter (CDM) cosmology N-body simulations, the structure of
dark halos, in particular the density profile of dark halos. Their
numerical simulations in the $\Lambda$-CDM scenarios led to the
density profile of galaxies as
\begin{equation}\label{E:rho}
\rho(r)=\frac{\rho_s}{(\frac{r}{r_s})(1+\frac{r}{r_s})^2}
\end{equation}
where $r_s$ is the characteristic scale radius and $\rho_s$  the
corresponding density. This density profile of CDM halos fits
accurately up to masses between $3 \times 10^{11} M_\odot  -  3
\times 10^{15} M_\odot$.

\subsection{The Universal Rotation Curve dark matter profile}

According to the NFW model, the velocities in the central parts
are too low~\citep{Gentile2004}, but in the outer regions of the
halo it fits well. In this case we consider with the range from
closer to the center to the outer region where the URC dark matter
profile~\citep{Castignani2012} is valid and given by
\begin{equation}\label{E:rho}
\rho(r)=\frac{\rho_0 r_0^3}{(r+r_0)(r^2+r_0^2)},
\end{equation}
where $r_0$ is the core radius and $\rho_0$ is the effective core
density.

The next ingredient is rotation curve regarding which it is
believed that rotation curve analysis is one of the great support
for the existence of dark matter in galaxies. By considering the $
H_{\alpha} $ data and also by adopting some radio rotation
curves~\citet{Persic1996} have analyzed a large number of rotation
curves and have argued that rotation curves can be fitted not only
for any luminosity, but also for any type of galaxies (may be
spirals, low-surface-brightness ellipticals and dwarf-irregular
galaxies). Therefore they used the term universal rotation curve
(URC) in stead of rotation curves.

The tangential velocity~\citep{Chandrasekhar1983} can be found
from the flat rotation curve for the circular stable geodesic
motion in the equatorial plane as
\begin{equation}\label{E:v1}
(v^{\phi})^2= r   f'(r).
\end{equation}

The velocity $v^\phi$ in km/s of the rotation curve of the objects
with total virial mass $3\times10^{12}$ of solar masses in
different radii $r$ in kpc is given in Table
1~\citep{Persic1996,Salucci2007}. By using interpolation
technique~\citep{Scarborough1966} for the data given in Table 1,
we have estimated fifth degree polynomial which is the best
fitting curve that yields the expression for the velocity as a
function of the radial coordinate $r$. Therefore the output can be
presented as
\begin{equation}\label{E:v2}
   v^{\phi}=0.0000011 r^5-.0003r^4+.031r^3 -1.4 r^2+28 r+64.
\end{equation}

One can note that the velocity curve we obtained fits well with
the observed tangential velocities in different distances (see
Fig. 1).

Here, the velocity $v^\phi$ is given in km/s which is equivalent
to~\citep{Rahaman2008,Nandi2009}
\begin{equation}\label{E:v2}
   v^{\phi}= (0.000003) (0.0000011 r^5-.0003r^4+.031r^3 -1.4 r^2+28
   r+64).
\end{equation}

\begin{figure}[h]
\centering
\includegraphics[scale=.4]{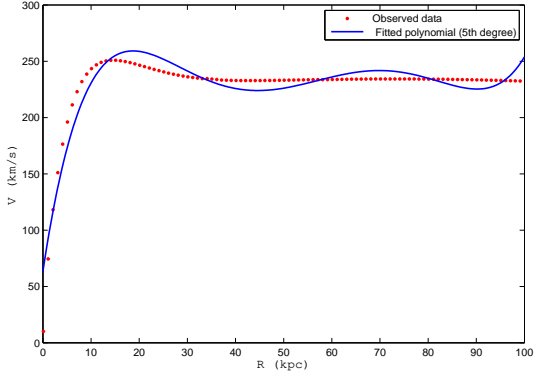} \caption{The
rotational velocity from observational data (Table 1) and
estimated fifth degree polynomial. The horizontal and vertical
lines are  distance $r$ in kpc and the rotation velocity $v_\phi$
in km/sec respectively.} \label{fig:8}
\end{figure}

\section{The Einstein field equations}

Now, the Einstein field equations for the above metric
(\ref{E:line1}) are
\begin{equation}
\frac{b^{\prime }(r)}{ r^{2}}=8\pi\rho (r),\label{E:Ein1}
\end{equation}

\begin{equation}
 2\left(1-\frac{b}{r}\right) \frac{f^\prime}{r}
 -\frac{b}{r^{3}}=8\pi p_r(r),\label{E:Ein2}
\end{equation}

\begin{eqnarray}
\left(1-\frac{b}{r}\right)\left[ f^{\prime \prime}+
\frac{f^\prime}{r} +{f^\prime}^2 - \left\{ \frac{b^{\prime
}r-b}{2r(r-b)}\right\}\left(f^{\prime} + \frac{1}{r}\right)\right]
\nonumber \\=8\pi p_{t}(r).\label{E:Ein3}
\end{eqnarray}

From Eqs. (\ref{E:v1}) and (\ref{E:v2}), one can get the
expression for redshift function $f$ as
\begin{eqnarray}\label{E:redshift}
f(r)  =(0.000003)^2 [1.21\times10^{-13}r^{10} -
7.33\times10^{-11}r^9 \nonumber \\ + 1.96\times10^{-8}r^{8}  -
3.1\times10^{-6}r^{7} + 3.1\times10^{-4}r^{6} - 0.0207r^{5}
\nonumber \\ + 0.914r^{4} - 24.81r^{3} + 302.4r^{2} + 3584 r +
4096 \ln  r].
\end{eqnarray}

The solution of the same redshift function will be used to obtain
the parameters from the Einstein field equations for different
cases. The logic behind this is that this redshift function is
obtained from the observed rotational velocities of the different
distance of outer regions of galaxies.

We note that our solutions valid up from $0.1$~kpc to $100$~kpc.
At the maximum distance of $100$~kpc, the redshift function does
not approach to zero. This means wormhole spacetime considered
here is not asymptotically flat and therefore $100$~kpc is the
{\it cut off} radius where these solutions will be joined smoothly
to an exterior vacuum solutions. Also we note that the redshift
function is finite in the range $0.1~ kpc < r < 100~ kpc$. This
finiteness of the redshift function actually prevents the event
horizon.

Now in the following, we will discuss two cases separately: For
the outer regions of the halo with $ 30~ kpc \leq r \leq 100 ~kpc$
({\it Case 1}) whereas for the central region with $ 9 ~ kpc \leq
r \leq 30~kpc$ ({\it Case 2}).

\subsection{Case 1}

The Einstein field equation yields the shape function as
\begin{equation}
b(r)  =8\pi \rho_s
r_s^3\left[\ln(r+r_s)+\frac{r_s}{r+r_s}\right]+C.
\end{equation}
where $C$ is an integration constant.

After obtaining both $f(r)$ and $b(r)$, we now examine whether the
spacetime possesses wormhole like geometry. From above, one can
see that \emph{redshift function} ($f(r)$) remains finite to
prevent an event horizon. We assume that characteristic scale
radius $r_s$ coincides with the throat radius of the wormhole,
such that  $b(r_s) = r_s$. This gives the value of the integration
constant $C$ as
\begin{equation}  C  =r_s - 8\pi \rho_s
r_s^3\left[\ln(2r_s)+\frac{1}{2}\right].
\end{equation}

Now, we find $b^\prime(r_s) < 1$, to check the so-called {\it
flare-out} condition. For the values  $\rho_s =0.0000001$ and $r_s
=30 ~kpc $, we obtain that $b^\prime(30) \approx 2.25 \times
10^{-4} < 1$. Thus the flare-out condition holds good. One can
note that this result does not modify if we implement the accurate
values of Milky Way for $\rho_s$ and $r_s$ found out
by~\citet{Nesti2013}.

The radial and lateral pressures assume the following forms:
\begin{eqnarray}
p_r(r) = \frac{1}{4\pi}\left[1-\frac{8\pi \rho_s
r_s^3(\ln(r+r_s)+\frac{r_s}{r+r_s}+C)}{r}\right] \nonumber \\
\times\left(\frac{f^{\prime}}{r}\right) -\frac{8\pi \rho_s
r_s^3(\ln(r+r_s)+\frac{r_s}{r+r_s}+C)}{8\pi r^3},
\end{eqnarray}

\begin{eqnarray}
p_t(r)= \frac{1}{8\pi}\left(1-\frac{8\pi \rho_s
r_s^3(\ln(r+r_s)+\frac{r_s}{r+r_s}+C)}{r}\right) \nonumber \\
\times \left[ f^{''}+\frac{f^{'}}{r}+(f^{'})^2 -\left\{
\frac{b^{\prime
}r-b}{2r(r-b)}\right\}\left(f^{\prime}+\frac{1}{r}\right)\right],
\end{eqnarray}
where $f(r)$ and $b(r)$ are given in Eqs. (11) and (12)
respectively and
\begin{eqnarray}
f^{\prime} =(0.000003)^2 [1.21\times10^{-12}r^8 -
6.6\times10^{-10}r^7 \nonumber \\ + 1.58\times10^{-7} r^6 -
2.168\times 10^{-5}r^5 + 1.86\times10^{-3}r^4 \nonumber \\ -
0.1035r^3 + 3.658r^2-74.43r + 604.8 \nonumber \\ + \frac{3584}{r}
+ \frac{4096}{r^2}],
\end{eqnarray}

\begin{eqnarray}
f^{''}=(0.000003)^2 [10.8\times10^{-12}r^8 - 52.8\times10^{-10}r^7
\nonumber \\ + 11.1\times10^{-7}r^6 - 13.02\times10^{-5}r^5 +
9.3\times10^{-3}r^4 \nonumber \\ - 0.414r^3 + 10.98r^2 - 148.86r +
604.8 - \frac{4096}{r^2}].
\end{eqnarray}

Fig. 2 indicates that $(\rho+p_r)<0$ and therefore the null energy
condition is violated to hold a  wormhole open.

\begin{figure}[h]
\centering
\includegraphics[scale=.3]{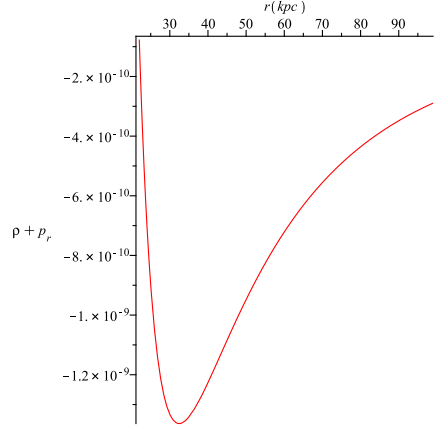}
\caption{The variation of the left-hand side of the expression for
the null energy condition of matter in the galactic halo is
plotted against r. } \label{fig:8}
\end{figure}

\subsection{Case 2}

In this case the shape function can be found as
\begin{equation}\label{E:redshift}
b(r)=8 \pi \rho_0
r_0^3\left[\frac{\log(r+r_0)}{2}+\frac{\log(r^2+r_0^2)}{4}
-\frac{tan^{-1}(\frac{r}{r_0})}{2}\right] +D
\end{equation}
where $D$ is an integration constant.

The radial and lateral pressures given by
\begin{eqnarray}
p_r(r)=~~~~~~~~~~~~~~~~~~~~~~~~~~~~~~~~~~~~~~~~~~~~~~~~~~~~~~~~~~~~~~~~~~~~~\nonumber
\\
 \frac{1}{4\pi}\left(1-\frac{8 \pi \rho_0 r_0^3\left(\frac{\log(r+r_0)}{2}+\frac{\log(r^2+r_0^2)}{4}
-\frac{tan^{-1}(\frac{r}{r_0})}{2}\right)+D}{r}\right) \nonumber
\\ \times
\left(\frac{f^{\prime}}{r}\right) -\frac{8 \pi \rho_0
r_0^3\left(\frac{\log(r+r_0)}{2}+\frac{\log(r^2+r_0^2)}{4}
-\frac{tan^{-1}(\frac{r}{r_0})}{2}\right)+D}{8\pi r^3},
\end{eqnarray}

\begin{eqnarray}
p_t(r)=~~~~~~~~~~~~~~~~~~~~~~~~~~~~~~~~~~~~~~~~~~~~~~~~~~~~~~~~~~~~~~~~~~~~~\nonumber
\\ \frac{1}{8\pi}\left[1-\frac{8 \pi \rho_0
r_0^3\left(\frac{\log(r+r_0)}{2}+\frac{\log(r^2+r_0^2)}{4}
-\frac{tan^{-1}(\frac{r}{r_0})}{2}\right)+D}{r}\right] \nonumber
\\  \times \left[f^{''}
+\frac{f^{'}}{r}+ {f^{'}}^2-\left\{\frac{b^{\prime
}r-b}{2r(r-b)}\right\}\left(f^{\prime}+\frac{1}{r}\right)\right],
\end{eqnarray}
where $f(r)$ and $b(r)$ are given in Eqs. (11) and (18).

One can check that the qualitative features meet all the
requirements for the existence of a wormhole based on the URC
model. Therefore, it would be desirable to examine the solutions
critically. We assume that the throat of the wormhole coincides
with the core radius $r=r_0$. As before, $b(r_0)=r_0$ yields the
value of the integration constant $D$ where
\begin{equation}
D = r_0 -4\pi r_0^3\rho_0 \ln 2r_0 - 2\pi r_0^3\rho_0 \ln 2r_0^2 +
\pi^2 \rho_0r_0^3.
\end{equation}

The known throat radius permits a closer look on the flare-out
condition, $b'(r_0) < 1$. Using the observed values of the Milky
Way galaxy, $r_0=9.11$~kpc~\cite{Maccio2012} and $\rho_0=5\times
10^{-24}(r_0/8.6 $~kpc)$^{-1}$ g~cm $^{-3}$~\cite{Castignani2012},
we get \[b'(r_0)\approx 1.74\times 10^{-6}<1.\]

Therefore, shape function obeys the flare-out condition. As
before, Fig. 3 indicates that $(\rho+p_r)<0$, therefore the null
energy condition is violated to hold a wormhole open.

\begin{figure}[h]
\centering
\includegraphics[scale=.3]{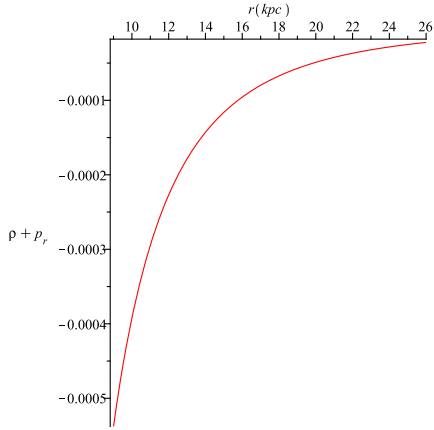}
\caption{The variation of the left-hand side of the expression for
the null energy condition of matter in the galactic halo is
plotted against r.} \label{fig:8}
\end{figure}

\section{Conclusion}
Recently, based on Navarro-Frenk-White (NFW)~\citep{Navarro1996}
density profile and the Universal Rotation Curve
(URC)~\citep{Castignani2012} dark matter
model,~\citet{Rahaman2014a,Rahaman2014b} have shown that the
galactic halo possesses the necessary properties for supporting
traversable wormholes. The former is valid for outer region of the
galaxies whereas the latter is valid for the central parts of the
galactic halo. For both the studies, they used the {\it ad hoc}
functional forms of rotation velocity.

However, in this paper by using the experimental data of the
rotational velocities at different distance of outer regions of
galaxies, we have estimated fifth degree polynomial that yields
the expression for the velocity as a function of the radial
coordinate $r$. This rotational velocity is used to find the
geometry of galactic halo regions within the framework of general
theory of relativity. The estimated rotational velocity function
$v_\phi$ is well behaved within the range $~0.1 ~kpc  ~\leq ~r~
\leq ~100~ kpc$. We have used this rotational velocity function
$v_\phi$ to find the spacetime geometries of outer region as well
as central parts of the galactic halo. Basically, we have in hand
the data of galactic rotational velocities from $~0.1 ~kpc~ to
~100 ~kpc $ and we have estimated fifth degree polynomial which is
the best fitted curve for velocity within this range. Therefore,
from our model we cannot predict what happens in galactic centre.
Also the wormhole exists outside the core of the galactic  halo.
Following~\citet{Maccio2012} we have assumed core radius as 9.11
kpc. One can define the central region as 9.11 kpc up to 30 kpc
and outer region as above 30 kpc. May be it needs further research
to predict the event in the galactic centre.

In this respect it is to note that we do not know whether it is
possible to explain the results without postulating the existence
of wormholes, rather by attributing some specific physical
properties to dark matter. However, according to our observations,
the dark matter in the galactic halo region produces the spacetime
geometry which is very similar to wormhole like geometry.

The present study provides a clue for  possible existence of
wormholes in most of the galaxies and provides a theoretical
platform to seek observational evidence for wormholes by studying
the scattering of scaler waves or from past data using ordinary
light as well as one can use the method of gravitational lensing
as a possible experiment~\citep{Kuhfittig2014}. Another suggestion
have given by~\citet{Torres1998} that wormholes can be probed
using light curves of gamma-ray bursts.

\section*{Acknowledgments}

FR and SR are grateful to the Inter-University Centre for
Astronomy and Astrophysics (IUCAA), India for providing
Associateship Programme. FR is thankful to the DST, Govt. of India
for financial support under PURSE programme whereas BS is thankful
to the DST for providing financial support under INSPIRE
programme. We are grateful to the referee for his valuable
comments which have improved the manuscript substantially.

\newpage

\begin{table}
\caption{The radius $r$ in kpc and velocity $v^\phi$ in km/s of the
Rotation Curve of objects with total virial mass $3\times 10^{12}$
solar masses. } \label{tab3} \centering
\bigskip
{\small

\begin{tabular}{cccccc}
\hline \\[-9pt]

$R$ (kpc) & $v^\phi$ (km/s) &~ $R$ (kpc) & $v^\phi$ (km/s) & $R$ (kpc) & $v^\phi$ (km/s)   \\

\hline \\
  0.1 &10.053 &34.1& 234.623
&68.1& 234.293
 \\

  1.1 &74.467 &35.1& 234.225
&69.1& 234.317\\

  2.1 &118.223 &36.1&233.891
&70.1& 234.334 \\

 3.1 & 151.113 &37.1&233.390
&71.1& 234.345\\

  4.1 &176.445 &38.1& 233.213
&72.1& 234.349\\

  5.1 &196.099 &39.1& 233.079
&73.1& 234.347\\

  6.1 &211.331 &40.1&232.982
&74.1& 234.339\\

  7.1 &223.057 &41.1&  232.918
&75.1& 234.324\\

  8.1 &231.975 &42.1& 232.883
&76.1& 234.303\\

  9.1 &238.634 &43.1& 232.873
&77.1& 234.303\\

  10.1 &243.475 &44.1& 232.884
&78.1& 234.276\\

  11.1 &246.857 &45.1& 232.913
&79.1& 234.243\\

  12.1 &249.072 &46.1& 232.957
&80.1& 234.205\\

 13.1 &250.362 &47.1& 233.013
&81.1& 234.160\\

  14.1 &250.927 &48.1& 233.078
&82.1& 234.109\\

  15.1 &250.930 &49.1& 233.151
&83.1& 234.054\\

  16.1 &250.509 &50.1& 233.230
&84.1& 233.993\\

  17.1 &249.774 &51.1& 233.313
&85.1& 233.927\\

  18.1 &248.817 &52.1& 233.397
&86.1& 233.856\\

  19.1 &247.712 &53.1& 233.483
&87.1& 233.779\\

  20.1 &246.519 &54.1& 233.568
&88.1& 233.698\\

  21.1 &245.285 &55.1& 233.652
&89.1& 233.612\\

  22.1 & 244.048 &56.1& 233.733
&90.1& 233.522\\

  23.1 &242.836 &57.1& 233.811
&91.1& 233.427\\

  24.1 &241.671 &58.1& 233.733
&92.1& 233.328\\

  25.1 &240.568 &59.1& 233.811
&93.1& 233.225\\

  26.1 &239.539 &60.1& 233.886
&94.1& 233.118\\

 27.1 &238.589 &61.1& 233.956
&95.1& 233.007\\

  28.1 &237.724 &62.1& 234.021
&96.1& 232.892\\

  29.1 &236.943 &63.1& 234.081
&97.1& 232.774\\

  30.1 &236.245 &64.1& 234.136
&98.1& 232.652\\

  31.1 &235.628 &65.1& 234.184
&99.1& 232.527\\

  32.1 & 235.089 &66.1& 234.227
\\

  33.1 &234.623 &67.1& 234.263
\\

\hline
\end{tabular}   }
\end{table}

\end{document}